\documentstyle[12pt,aasms4]{article}

\def\alwaysmath#1{\ifmmode{#1}\else{$#1$}\fi}
\def\msun{\alwaysmath{\,{M}_{\odot}}}

\def\logteff{\alwaysmath{\log T_{\rm eff}}}
\def\logl{\alwaysmath{\log L}}
\def\feh{{\rm [Fe/H]}}
\def\ofe{{\rm [O/Fe]}}

\def\ebmv{\alwaysmath{E(B-V)}}
\def\mcore{\alwaysmath{M_{\rm core}}}

\def\etal{{et~al.\ }}
\def\SP{$2^{\rm nd}$P}
\received{1 Mar 1999}
\accepted{18 May 1999, ApJLetters}
\lefthead{Majewski et al.}
\righthead{Second Parameter in the Sculptor Dwarf}

\begin{document}
\title{An Internal Second Parameter Problem in the Sculptor Dwarf
  Spheroidal Galaxy}

\author{S. R. Majewski\altaffilmark{1,2}, M. H. Siegel\altaffilmark{1},
 Richard J. Patterson\altaffilmark{1}, and R. T. Rood\altaffilmark{1}
}

\altaffiltext{1}{Department of Astronomy, University of Virginia, P.O.
  Box 3818, Charlottesville, VA, 22903-0818
  (srm4n@didjeridu.astro.virginia.edu, mhs4p@virginia.edu,
  ricky@virginia.edu, rtr@virginia.edu)}

\altaffiltext{2}{Visiting Research Associate, The Observatories of the
  Carnegie Institution of Washington, 813 Santa Barbara Street,
  Pasadena, CA 91101; David and Lucile Packard Foundation Fellow;
  Cottrell Scholar of the Research Corporation.}

\begin{abstract}

  We present $BV$ photometry of the Sculptor dwarf galaxy to $V=22$.
  These data give evidence for a bimodality in Sculptor's metallicity
  distribution based on a discontinuity in the luminosities of
  horizontal branch (HB) stars and by the presence of two distinct red
  giant branch (RGB) bumps.  A consistent picture of the evolved stars
  in Sculptor is given by the presence of (1) a metal-poor population of
  $\feh \sim -2.3$ with an exclusively blue HB and that corresponds to
  the blueward side of the Sculptor RGB and the more luminous RGB bump,
  and (2) a less metal-poor population of $\feh \sim -1.5$ required to
  explain the less luminous red HB, the red side of the RGB, and a
  second, less luminous RGB bump.  Best fits to the HB
  populations are obtained with enhanced oxygen abundances, $\ofe \sim
  +0.5$.  Variations in the global HB and RGB morphology of Sculptor can
  be explained by differences in the radial distribution of these two
  populations.  The presence of these two populations shows that the
  Sculptor dwarf galaxy has an {\it internal} second parameter problem.

\end{abstract}


\clearpage

\section{Introduction}

The Sculptor dwarf spheroidal (dSph) galaxy was the first Galactic dSph
to be identified (Shapley 1938) and has a long observational history.
Its variables have been tabulated by a large number of studies (Baade \&
Hubble 1939; Thackeray 1950; van Agt 1978; Goldsmith 1993, Hereafter
G93; Ka{\l}u\`zny \etal 1995, hereafter K95), and the period
distribution of RR~Lyrae stars suggests a metallicity spread (G93, K95).
While a range of abundances is generally accepted, impressions of the
Sculptor horizontal branch (HB) morphology outside of the instability
strip have varied depending on photometric depth, field size and filter
systems employed to construct the color magnitude diagram (CMD).  Kunkel
\& Demers' (1977, KD77) Sculptor CMD (324 stars to $V=20.6$) yielded 43
red HB (RHB) out of 49 HB stars and a deficit of stars with $B-V<+0.3$
as well as a red giant branch (RGB) well described by a metal-poor
population ($\feh =-1.9$).  Norris \& Bessell (1978) re-analyzed the CMD
in combination with two spectra to argue for a Sculptor metallicity
spread of $-2.2 \le \feh \le -1.5$, and Smith \& Dopita (1983) confirmed
an inhomogeneous metallicity distribution function (MDF) via narrow-band
photometry.  Da Costa's (1984, D84) deep, but small area photometry to
the Sculptor MSTO did not provide strong constraints on either the HB or
RGB; however, it did show an abundance spread similar to previous
results (confirmed by Da Costa 1988) and a predominantly {\it red} HB (7
of 10 HB stars).  The conclusion of these studies was that Sculptor is a
``second parameter (\SP) object'' that shows a rather red global HB for
its mean abundance (D84).

More recently, however, Schweitzer \etal (1995, SCMS) produced a CMD
with 1043 stars that reveals a prominent {\it blue} HB (BHB) with more
stars than KD77 and D84.  In the first wide-field, {\it CCD} survey of
Sculptor, K95 reported the usual metallicity spread based on the RGB (as
did SCMS), but also substantiated the large BHB population and derived a
more moderate Sculptor HB morphology index of $(B-R)/(B+V+R) = -0.15$
(see also the Grebel \etal 1994 CMD).  Because their $VI$ photometry of
$>6000$ stars with $V<21.5$ covered a much larger area than previous
results, K95's significantly increased BHB:RHB ratio suggests
differences in the spatial distribution of BHB and RHB stars.  This
might be due to abundance gradients in the dSph. However, since the most
metal rich population in Sculptor's RGB has $\feh \sim -1.5$, which
would normally give a uniform to blue HB, the variation in the RHB
population must be due to spatial variation in the {\it \SP} effect.

In this letter, we present $BV$ photometry of the Sculptor dSph galaxy.
We find the usual evidence for RGB stars ranging from metal-poor ($\feh
\sim -1.5$) to very metal-poor ($\feh \sim -2.3$); however, on the basis
of two {\it distinct} HBs and two distinct RGB bumps, Sculptor's MDF may
be better characterized as {\it bimodal}.  This bimodality gives rise to
one population with a \SP\ effect, and a second one with likely very
little HB \SP.  Differences in radial distributions for these two
populations can account for the variation in HB morphology within
Sculptor and among previous surveys of this galaxy.
 
\section{Observations and Reduction}

We observed Sculptor on UT 23 July and 1--2 August, 1991 with the Las
Campanas 1-m Swope telescope using the thinned, $1024^2$ TEK2 CCD
camera.  Five overlapping, 10\farcm4 wide pointings were arranged in a
2$\times$2 grid with a center frame overlapping the other four to lock
together the photometry.  Each field was typically observed with one $B$
and $V$ exposure of 1800 and 900 sec length, respectively.  The data
were reduced with the IRAF package CCDRED and photometered with the
DAOPHOT II and ALLFRAME programs (Stetson 1987, 1994).  Detections were
matched using DAOMASTER and then calibrated to observed Graham (1982)
standard stars using our own code.  This code compares calibrated
magnitudes of stars in common on different CCD frames and determines
minor frame-to-frame systematic errors (e.g., due to shuttering errors,
transient transparency changes, errors in the photometric
transformation).  Because of photometric conditions, the derived mean
residuals for each frame ($\le 0.1$ mag on the basis of $\ge689$
comparison stars) were used as offsets and applied iteratively with new
color determinations until convergence.  Our resulting photometric
precision is $(\sigma_{B},\sigma_{V})=(0.05,0.05)$ mag at the HB.

\section{Horizontal Branch}

Our $(B-V,~V)_0$ (Figure 1) and $(B-V,~B)_0$ (not shown) CMDs show an HB
that appears to be kinked over the RR~Lyrae gap.  All tests of the
photometry pipeline have shown this ``kink'' to be real, and a hint of
this HB ``kink'' can be seen in KD77.  Similarly kinked HBs have been
noted previously in the CMDs of some ``bimodal'' Galactic globular
clusters (GGC), e.g., NGC~6229 (Borissova \etal 1997), NGC~2808 (Ferraro
\etal 1990), and NGC~1851 (Walker 1992), to which the Sculptor CMD bears
some resemblance.  Indeed, our derived $B:V:R$ (blue:variable:red HB)
ratio of (0.42:0.19:0.39) resembles those of bimodal GGCs (see Borissova
\etal 1997 for a summary).  Stetson \etal (1996) make a poignant
comparison of the bimodal NGC~1851 CMD to those of the similar
metallicity, ``\SP\ GGC pair'' NGC~288 and NGC~362; that NGC~1851 has
{\it both} an RHB like NGC~362 {\it and} a BHB like NGC~288 suggests
that NGC~1851 has an internal \SP\ problem.  Stetson \etal use this fact
to argue as unlikely that the \SP\ effect is due to differences in age,
helium abundance or [CNO/Fe] within NGC~1851.  Despite the similarities
of our Sculptor HB to the HBs of bimodal GGCs, there are two key reasons
why the Stetson \etal analysis does not apply here: (1) From the RGB
width we {\it know} that Sculptor has an abundance spread. (2) There is
no {\it a priori} reason to assume that all of the stars in Sculptor are
coeval. Bearing this in mind we now explore the origin of the bimodality
of the Sculptor HB.

The $V$ magnitude difference from
the red edge of the Sculptor BHB to the blue edge of the RHB is 
$\sim0.15 \pm 0.02$ mag.  If the bimodality is completely
due to differences in [Fe/H], typical values for
$dM_V/d{\rm \feh}$ suggest 
an [Fe/H] difference of 0.5 to 1.0 dex.  This is
consistent with reported [Fe/H] spreads from fitting isochrones
to the Sculptor RGB. 

The situation is, however, more complex because we are comparing HB
stars at different colors, and both the luminosity of the theoretical
ZAHB (zero-age-HB) and bolometric correction vary with position along
the HB. Moreover, the HB is strongly affected by the oxygen
abundance. At a constant core mass (\mcore) increasing [O/Fe]
increases $L_{\rm HB}$. However increasing [O/Fe] also leads to a
decrease in \mcore.  All other things being equal, a decrease in \mcore\
leads to a decrease in $L_{\rm HB}$. The net result is that the ZAHB
variation with [O/Fe] can be rather complex.  In the Galaxy it is
generally thought that for metallicities appropriate for Sculptor,
[O/Fe] is constant with a value
in the range +0.3--0.5. There is no reason to assume that Sculptor has
undergone the same chemical enrichment history as the Galaxy so we
consider all $ 0.0 \le \ofe \le +0.5$ possible.  Most recent HB models
have an assumed \ofe, \feh\ relation.  The only available models that 
allow us to explore the composition parameters independently are those of
Rood (unpublished).  To convert \logl\ and \logteff\ to $M_V$ and
$B-V$ we use the results of Kurucz (1979) and Bell \& Gustaffson
(1978) blended to reproduce observed HBs of GGCs smoothly.  Throughout
this paper, we assume $(m-M)_0 = 19.71$ and $\ebmv = 0.02$ (K95) for
Sculptor.

Figure 2 shows the observed CMD of the Sculptor HB with superimposed
ZAHBs terminated at the red end at a mass of 0.85\msun with a cross mark
indicating a mass of 0.80\msun. These are roughly the maximum possible
masses for 12 and 15 Gyr populations, respectively.  Since all stars
undergo some mass loss the ZAHB population will not actually reach these
two points. Evolution and observational scatter will carry some stars
redward, but for practical purposes the end of the ZAHB
should mark the redward extent of the HB.

We start with the hypothesis that the Sculptor BHB is a low
metallicity population and the RHB a higher metallicity population,
both consistent with the spread of the RGB. The fairly uniform
distribution across the RGB suggests comparable numbers in each
group. The size of the observational error would obscure obvious
bimodality on the RGB. 

The BHB can be fit reasonably with $\feh = -2.3$ and $0.0 \le \ofe \le
+0.5$.  Indeed, the BHB rather resembles that of the low metallicity
GGC M92 (see Figure 1). The RHB can be fit with oxygen enhanced models
with $-1.9 \le \feh \le -1.5$. The odd behavior of the ZAHB level with
\feh\ for the $\ofe = +0.5$ ZAHBs is due to approximations used for
\mcore. Independently of such modeling details, one can expect the
variation of ZAHB level with \feh\ to be less for oxygen enhanced
models than for scaled solar abundances. The models with $\ofe = 0.0$
cannot fit the RHB: at $\feh = -1.9$ the ZAHB does not extend far
enough to the red; at $\feh = -1.5$ the level of the ZAHB is too low.

One could conceivably produce the observed bimodality using one
composition with $-1.9 \le \feh \le -1.5$ and $\ofe=+0.5$. Such a
mono-compositional bimodality is observed in GGCs but modeling it
requires the {\it ad hoc} introduction of bimodality in some underlying
parameter (Catelan \etal 1998). However, in Sculptor a composition
spread is observed, and a bimodal composition is quite natural, e.g.,
arising from two bursts of star formation. Hence, it seems undesirable
to us to discard the ``natural explanation'' in favor of the yet to be
determined mechanism that produces bimodal HBs in GGCs.

It is clear from the $\ofe = +0.5$ ZAHB (Figure 2a) that even if most of
the low metallicity population is found on the BHB, some could be
found in the RR~Lyrae strip and on the RHB. We suspect that this is
a small fraction of the low metallicity population, because the red end of
the BHB veers away from the ZAHB suggesting that the ZAHB is populated
only for $(B-V)_0 \lesssim 0.15$. In analogy to M92 we suspect that
$\ga 90$\% of the low metallicity population is found on the BHB and
that its age is similar to that of M92.

Similarly, from Figure 2 we see that the higher metallicity
population could contaminate the BHB.  The RHB
population does drop as one approaches the RR~Lyrae strip. But there is
precedent from the bimodal HB GGCs that such a population could
increase further to the blue. There is reason to think this is not
true for Sculptor. First, if there is significant high metallicity
contamination of the BHB, where are the low metallicity stars we infer
must be present from the RGB spread? Second, the BHB morphology is
more like that of M92 than the blue-HB-tails of clusters with higher
metallicity--M13, NGC~288, etc. These arguments in themselves are not
compelling, but fit the overall scenario we develop here.

Normal GGCs with the metallicity we suggest for the Sculptor RHB have
uniform HBs. This means that the Sculptor high metallicity population
suffers from a ``too red'' \SP\ problem like, e.g., GGCs NGC~362 and
NGC~7006 and the extreme halo cluster Pal 14. While the case for age as
the \SP\ in GGCs has been hard to establish (e.g., Stetson \etal 1996;
Catelan \etal 1998; VandenBerg 1998; but see counter views by Chaboyer
\etal 1996, Sarajedini \etal 1997), there is good reason to think that a
higher metallicity population in a low density system like Sculptor
might be younger. Thus, we hypothesize that the RHB arises from a
population several Gyr younger than the BHB.  Indeed, D84 has suggested
multiple age components ($\delta({\rm age}) \sim3$ Gyr) from his study
of turnoff stars.

If bimodal, Sculptor's two HB populations probably overlap significantly
in the instability strip.  The distribution of RR~Lyrae periods in
Sculptor (G93, K95) shows a large range, consistent with a large spread
in metallicity. The periods of RRab stars at the blue fundamental edge
of the instability strip (those with the shortest periods) are well
correlated with the metallicity (Sandage 1993a).  In Sculptor, the
shortest period RRab (ignoring two stars with very discrepant periods)
has a period of 0.474 days (K95), implying a metallicity of $\feh=-1.6$.
While the red fundamental edge is not as useful a metallicity indicator,
the existence of RRab stars with $P\gtrsim0.8$ days indicates the
presence of another population with $\feh<-2.0$.  In addition, G93 and
K95 both note a correlation of average magnitude with period in Sculptor
RRab stars. Because $\left<M_V\right>$ is a function of $\feh$, the
spread in $\left<M_V\right>$ also implies a metallicity spread. The
intensity weighted average $\left<V\right>$ magnitude for the majority
of the RRab stars lies in the range $20.1 < \left<V\right> < 20.25$
(K95), or $ 0.24 < \left<M_V\right> < 0.64$, which corresponds to $-2.3
< \feh < -1.3$ (Sandage 1993b).

\section{Red Giant Branch}

Our analysis so far points to a bimodality of populations in the
Sculptor HB.  However, such bimodality is also suggested in the giant
branch, where two distinct RGB bumps can be seen (Figure 1): one near
$(B-V, V)_0 = (0.8, 19.3)$ and one near $(0.8, 20.0)$.  The former RGB
bump lies toward the blue side of the RGB, near the expected locus for
metal poor stars, while the latter RGB bump lies toward the red side of
the RGB, near the expected locus for more metal rich stars.  To
illustrate the differences, we fit a mean RGB locus to the entire
Sculptor RGB, divide the RGB in half, and plot (Figure 3) RGB luminosity
functions for all stars within $\Delta (B-V) \sim 0.125$ left and right
of the mean RGB locus.  We isolate the redward RGB bump at $V_0 \sim
20.0$. The blueward bump is less clearly defined but probably is $19.0
\le V \le 19.4$. The extreme magnitude differences between the RGB bumps
again argues for a metallicity separation of order a dex.  We can use
the absolute magnitudes of the RGB bumps to obtain a global metallicity
([M/H]) for the two bump populations (Ferraro \etal 1999): Using
$(m-M)_0 = 19.71$ (K95), we find [M/H] $\sim -2.1$ and $\sim -1.3$.
Ferraro \etal (1999, Figure 11a) also give relations for the RGB bump
dependence on the magnitude difference between the bump and HB.  If we
adopt $V = 20.2$ for the height of the BHB population and assign this to
the metal-poor RGB, we obtain $V_{bump} - V_{HB} \sim -0.9$; this
implies an abundance $\feh \lesssim -2.4$, on the Zinn (1985) scale.
For the RHB population, if we adopt $V_{HB}=20.35$ and assign to this
the other RGB bump, we obtain $V_{bump} - V_{HB} = -0.35$; this is the
difference expected for $\feh \lesssim -1.6$.

The presence of the distinct RGB bumps, their estimated $M_V$, and
their location relative to the HB suggest a bimodal
MDF with $\feh \sim -2.3$ and $\feh \sim -1.6$.

\section{Discussion}

 From analysis of the RGB and HB, a consistent scenario can be
assembled. In Figure 1 we show representative RGB, AGB, and HB
fiducials for the metal-poor ($\feh=-2.23$), BHB cluster M92 and the
less metal poor ($\feh=-1.44$) \SP\ cluster Pal 14.  
These clusters bracket the Sculptor RGB, while each
cluster separately approximates the BHB and RHB, respectively.
Apart from the fact that Pal 14 may be a little metal-rich by a few
0.1 dex, the two clusters provide a reasonable
bimodal paradigm for the Sculptor MDF.

Our bimodal interpretation of Sculptor differs somewhat from previous
studies that argue for an abundance {\it spread}.  It should be noted
that a true bimodality in the RGB of Sculptor in the form of two
distinct RGB sequences would be masked somewhat by observational scatter
and the superposition of the asymptotic giant branch for the more metal
rich population.  The presence of two {\it distinct} RGB bumps, rather
than a slanting RGB bump ``continuum,'' is evidence for bimodality in
Sculptor.  We note that a suggestion of bimodality (or {\it
  tri}modality) was made previously by Grebel \etal (1994).

In \S 1 we argued that disparate HB morphologies found among different
surveys of the Sculptor CMD derived from radial differences in global HB
morphology.  Figure 4 provides evidence that this is the case: The
global HB index increases by 0.4 from the center to the $\sim 500"$
radius accessible with our catalogue.  We have argued for a {\it
  bimodal} MDF.  Accordingly, the radial gradient in Figure 4 is not
likely due to a radial abundance gradient, or the gradual diminishing of
a \SP\ effect.  Rather, the cumulative evidence suggests that the HB
radial dependence is due to changes in proportions of two nearly
mono-metallic populations with radius.  Indeed, the relative densities
of the blue:red half of the RGB track those of the BHB:RHB very well
(Figure 4).  The spatial distribution of the [BHB, blue RGB, metal-poor]
population appears to be more extended than that of the [RHB, red RGB,
less metal poor] population, which shows a higher core concentration.
Spatial differences in the Sculptor HB were suggested previously by
Light (1988) and Da Costa \etal (1996) and are explored further by
Hurley-Keller \etal (1999).  Da Costa \etal (1996) also point out radial
HB index gradients (with a similar sense) in the Leo II and And I dSphs,
and adopt the same interpretation of mixing variations in bimodal HB
populations.

The existence of bimodal, \SP\ $+$ non-\SP\ populations within dSphs
would be significant since, unlike bimodal GGCs such as NGC~1851, in
dSphs it is (now) entirely plausible to consider multiple star formation
bursts with age as the \SP.  In Sculptor's case, it is likely that the
$\feh \sim -2.3$ population formed in an earlier, more extended burst.
If the presence of these two distinct populations is born out, the
(relatively nearby) Sculptor dSph could well prove to be a Rosetta stone
of the HB and the adamantine \SP\ question.

We thank Eva Grebel for helpful discussions and the referee for useful suggestions.

\newpage
\section{Captions}

{\sc fig}. 1. \ -- \ $(B-V,~V)_0$ CMD for the Sculptor dwarf galaxy with
overlaid fiducials for M92 ({\it dashed line}; from Sandage, 1970) and
Pal 14 ({\it solid line}; from Holland \& Harris 1992).  We adopt
cluster distance moduli and reddenings from Harris (1996) and Holland \&
Harris (1992), respectively.  The right panel highlights the RGB and red
bump region.

{\sc fig}. 2. \ -- \ Fits of model ZAHBs to the HB of Sculptor.  Panel
(a) shows oxygen enhanced ($\ofe=+0.5$) models, and (b) shows models
with solar $\ofe$. In each panel, the {\it solid line} shows the model
with $\feh=-2.3$, the {\it dashed line} shows $\feh=-1.9$, and the {\it
  dotted line} shows $\feh=-1.5$.  The ZAHBs are terminated at the red
end at a mass of 0.85\msun and a cross mark indicates a mass of
0.80\msun (see text).

{\sc fig}. 3. \ -- \ Differential (right ordinate) and cumulative (left
ordinate) RGB luminosity functions for stars within 0.125 mag in
$(B-V)_0$ color to either the blue ({\it dot-dash curves}) or red ({\it
  solid curves}) of the mean RGB locus.  The {\it dot-dash curves} are
offset vertically by $+1.0$ for the cumulative and by $+80$ for the
differential luminosity function.  Breaks in the slope of the cumulative
distributions (indicated by {\it thin solid lines}) point to locations
of RGB bumps, marked by vertical lines.

{\sc fig}. 4. \ -- \ Radial dependence of HB ({\it filled circles}) and
RGB ({\it open circles}) morphology from our catalogue.  The RR~Lyrae
counts in the same areas are from K95. The values for $(B:V:R)$ for the
HB and $B+R$ for the RGB in each annulus are given for each point.
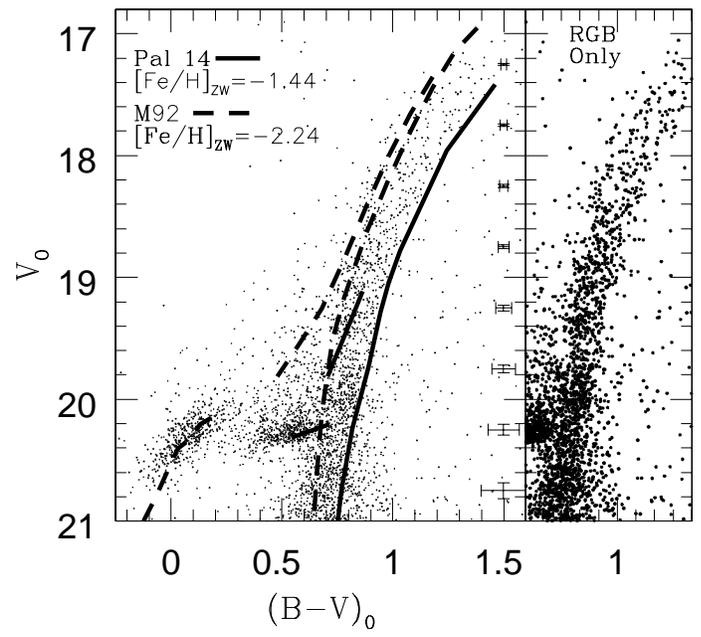
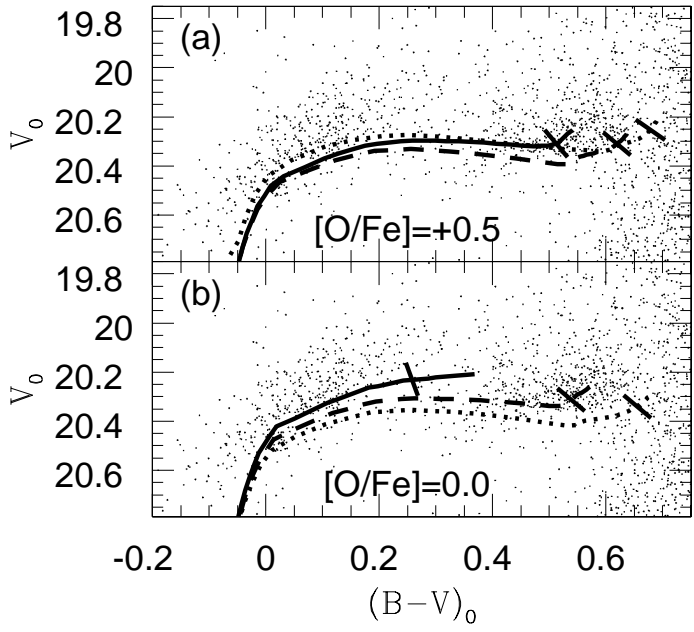
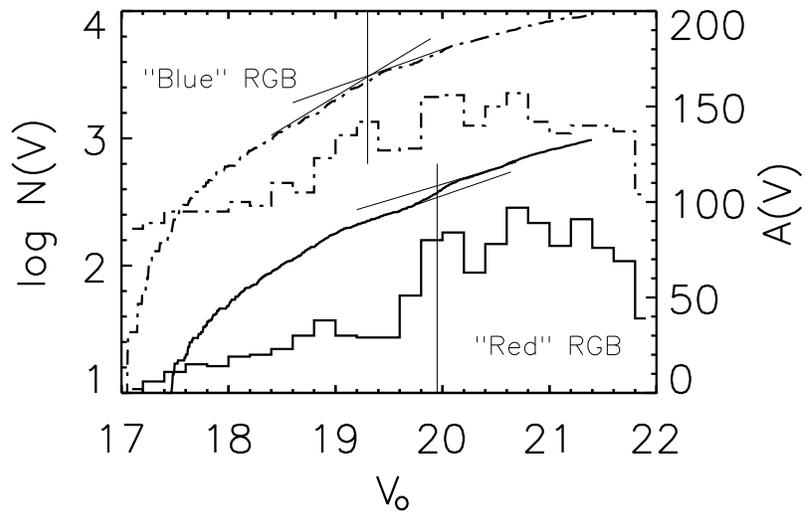
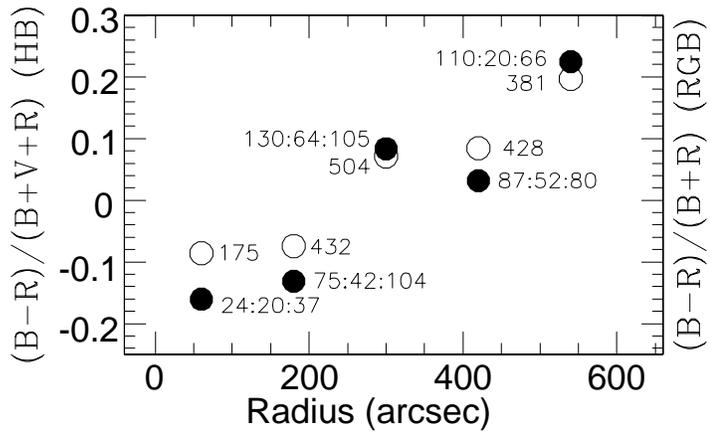
\end{document}